\documentclass[showpacs,10pt,twocolumn,prb]{revtex4-1}

\usepackage{amsmath}
\usepackage{amssymb}
\usepackage{graphicx}
\usepackage{amssymb}
\usepackage{graphics}
\usepackage{epsfig}
\usepackage{CJK}
\usepackage{color}

\begin{document}

\begin{CJK*}{GBK}{Song}
\title{Polaronic Conductivity in Cr$_2$Ge$_2$Te$_6$ Single Crystals}
\author{Yu Liu,$^{1}$ Myung-Geun Han,$^{1}$ Yongbin Lee,$^{2}$ Michael O. Ogunbunmi,$^{3}$ Qianheng Du,$^{1,4}$ Christie Nelson,$^{5}$ Zhixiang Hu,$^{1,4}$ Eli Stavitski,$^{5}$ David Graf,$^{6}$ Klaus Attenkofer,$^{5}$ Svilen Bobev,$^{3}$ Liqin Ke,$^{2}$ Yimei Zhu,$^{1}$ and C. Petrovic$^{1,4}$}
\affiliation{$^{1}$Condensed Matter Physics and Materials Science Department, Brookhaven National Laboratory, Upton, New York 11973, USA\\
$^{2}$Ames Laboratory, U.S. Department of Energy, Ames, Iowa 50011, USA\\
$^{3}$Department of Chemistry and Biochemistry, University of Delaware, Newark, Delaware 19716, USA\\
$^{4}$Materials Science and Chemical Engineering Department, Stony Brook University, Stony Brook, New York 11790, USA\\
$^{5}$National Synchrotron Light Source II, Brookhaven National Laboratory, Upton, New York 11973, USA\\
$^{6}$National High Magnetic Field Laboratory, Florida State University, Tallahassee, Florida 32306-4005, USA}
\date{\today}

\begin{abstract}
Intrinsic, two-dimensional (2D) ferromagnetic semiconductors are an important class of materials for spin-charge conversion applications. Cr$_2$Ge$_2$Te$_6$ retains long-range magnetic order in bilayer at cryogenic temperatures and shows complex magnetic interactions with considerable magnetic anisotropy. Here, we performed a series of structural, magnetic, X-ray scattering, electronic, thermal transport and first-principles calculation studies which reveal that localized electronic charge carriers in Cr$_2$Ge$_2$Te$_6$ are dressed by surrounding lattice and are involved in polaronic transport via hopping that is sensitive on details of magnetocrystalline anisotropy. This opens possibility for manipulation of charge transport in Cr$_2$Ge$_2$Te$_6$ - based devices by electron-phonon- and spin-orbit coupling-based tailoring of polaron properties.
\end{abstract}
\maketitle
\end{CJK*}

\section{INTRODUCTION}

Ferromagnetic semiconductors are of great importance for the newly-developing spintronics technology that allows for control of both charge and spin degrees of freedom of charge carriers.\cite{Wolf} The mechanism of intrinsic long-range magnetic order in monolayers of 2D materials such as CrI$_3$ and bilayer Cr$_2$Ge$_2$Te$_6$ might be exploited for design of novel spin-related devices.\cite{Huang,Gong} They can be stacked into heterostructures\cite{Alegria,Yao,Escolar,Mogi,Lix,ParkBG} and their physical properties are highly tunable by carrier doping,\cite{Tang,Tang1,Peng,Wang0} strain,\cite{Sivadas,Chittari,Zollner} vacancies,\cite{DY,Hao,Song0,Song,Milosavljevic} pressure\cite{Lin,Sun,Yu,Ge,Dong} and electric field.\cite{Xing,Chen,Wang,Mak}

Cr$_2$Ge$_2$Te$_6$ exhibits a paramagnetic (PM) to ferromagnetic (FM) transition with the Curie temperature ($T_c$) of 61 $\sim$ 68 K in bulk, an out-of-plane easy axis and negligible coercivity.\cite{Gong,Carteaux2} The $T_c$ is absent in monolayer but persists in bilayer Cr$_2$Ge$_2$Te$_6$ of a relatively high value of 30 K.\cite{Gong} Few-layer Cr$_2$Ge$_2$Te$_6$ crystals hold an intrinsic FM order with high tunability by gating, as required of a good functional quantum material in spintronics.\cite{Wang} Cr$_2$Ge$_2$Te$_6$ is proposed to be a two-dimensional (2D) Heisenberg ferromagnet based on the spin wave theory,\cite{Gong} but was also found to follow a 2D Ising-like or tricritical mean-field model,\cite{YL,LinGT} indicating a complex magnetic mechanism.

Cr$_2$Ge$_2$Te$_6$ has a semiconducting band gap of 0.2 $\sim$ 0.74 eV, which can be tuned into metallic by applying pressure, organic ion intercalation, and vacancies.\cite{Wang0,Song0,Yu} In a double-layer Cr$_2$Ge$_2$Te$_6$ device subjected to electrostatic gating metallic resistivity has been observed.\cite{Verzhbitskiy} Strong spin-phonon coupling has been verified in isostructural Cr$_2$Si$_2$Te$_6$.\cite{Casto} Yet, the electronic transport mechanism is still unclear in conducting Cr$_2$Ge$_2$Te$_6$ single crystals.

In this letter we show that local bond distances obtained from X-ray absorption spectroscopy exhibit little change when compared to published average values, implying little difference in the basic structural units of each atom. On the other hand, stacking of such units at longer atomic distances results in stacking faults. We observe relatively low values of thermal conductivity $\kappa$(T) below 300 K which is enhanced in magnetic field pointing to strong spin-phonon coupling. Moreover, electrical and thermal transport conduction reveal polaronic transport, commonly observed in oxide materials.\cite{Palstra} Due to strong spin-orbit coupling which is the source of magnetic anisotropy, polaronic transport mechanism could allow for tunable spin current in future spintronic devices.\cite{Covaci,Watanabe,Ostwal}

\section{RESULTS AND DISCUSSION}

The obtained single crystals exhibits a 2D layered characteristic with the cleave surface in the $\mathbf{ab}$-plane and the $\mathbf{c}$-axis is perpendicular to this surface (\textbf{Figure 1}a).\cite{Experimental} The lattice constant of about $c = 20.57(2)$ {\AA} can be extracted by the Bragg's law. Cr$_2$Ge$_2$Te$_6$ crystalizes in a layered structure with the $R\bar{3}h$ space group, in which the layers stack along the (00l) direction (\textbf{Figure 1}b,c). There are different types of chemical bonds in crystal structure, including strong intralayer Cr-Te ionic bonds, Ge-Te covalent bonds, Ge-Ge dimers, and weak interlayer van der Waals (vdW) forces.\cite{Ji} Each Cr is octahedrally surrounded by six Te, and the edge-sharing CrTe$_6$ octahedra form a honeycomb network. Dimers of Ge-Ge occupy the void of each honeycomb, forming Ge$_2$Te$_6$ groups.

\textbf{Figure 1}d shows the atomic resolution  high-angle annular dark-field (HAADF) scanning transmission electron microscopy (STEM) image along the (00l) direction. The Te atoms are hexagonally packed from the top view, while the Cr ions and Ge-Ge dimers occupy Te-octahedral sites. The corresponding electron diffraction pattern (inset in \textbf{Figure 1}d) confirms the space group. The chemical composition gives elemental ratio Cr : Ge : Te = $20.09:20.05:59.86$ (\textbf{Figure 1}e).\cite{Experimental} We found the presence of stacking faults in cross-section samples, as shown in \textbf{Figure 1}f-j. There are at least about 8 stacking faults in 300 nm length scale along the $c$-axis (\textbf{Figure 1}g-j), which gives 8/(300 nm) = 0.027 nm$^{-1}$ stacking fault density. This represents the upper limit since most of stacking faults are found near the original crystal surface.

The FM arises from the super-exchange between the nearest-neighbor (NN) Cr linked by Te ligands through nearly-90$^\circ$ angle, however the exchange interactions beyond the NN spins are also crucial in determining the magnetic ground state.\cite{Sivadas} Pressure can compress the bond lengths of Cr-Cr and Cr-Te and tilt the Cr-Te-Cr angle from 90$^\circ$, resulting in an enhancement of antiferromagnetic (AFM) interaction.\cite{Ji} This can also induce spin reorientation transition from an uniaxial to easy-plane anisotropy,\cite{Lin} indicating that the magnetism is strongly dependent on Cr local atomic environment. To probe this local environment we performed X-ray absorption near edge structure (XANES) and extended X-ray absorption fine structure (EXAFS) experiments. \textbf{Figure 2} shows the normalized Cr, Ge, and Te K-edge XANES spectra and the Fourier transform magnitudes of EXAFS spectra of Cr$_2$Ge$_2$Te$_6$, respectively. The XANES spectra (\textbf{Figure 2}a-c) are close to that of Cr$_2$Te$_3$ with Cr$^{3+}$ and Te$^{2+}$ states.\cite{Ofuchi} The local environment of indicated atoms is revealed in the EXAFS spectra of Cr$_2$Ge$_2$Te$_6$ measured at room temperature (\textbf{Figure 2}d-f). In a single-scattering approximation, the EXAFS can be described by:\cite{Prins}

\begin{align*}
\chi(k) = \sum_i\frac{N_iS_0^2}{kR_i^2}f_i(k,R_i)e^{-\frac{2R_i}{\lambda}}e^{-2k^2\sigma_i^2}sin[2kR_i+\delta_i(k)],
\end{align*}

where $N_i$ is the number of neighbouring atoms at a distance $R_i$ from the photoabsorbing atom. $S_0^2$ is the passive electrons reduction factor, $f_i(k, R_i)$ is the backscattering amplitude, $\lambda$ is the photoelectron mean free path, $\delta_i$ is the phase shift, and $\sigma_i^2$ is the correlated Debye-Waller factor measuring the mean square relative displacement of the photoabsorber-backscatter pairs. The main peaks have been corrected by standards of the NN Cr-Te [2.76(7) {\AA}] and Ge-Te [2.45(13) {\AA}] in the Fourier transform magnitudes of EXAFS. We summarized the local bond distances from different central atom ranging from 1.5 to 4.5 {\AA} in \cite{Experimental}. There are six Te NN for the Cr atomic site with two different distances [2.76(7) {\AA} and 2.87(7) {\AA}]; the next NN is one Cr [3.95(25) {\AA}], and then are six Ge with two different distances [4.11(25) {\AA} and 4.17(25) {\AA}]. Ge atomic site has three NN Te [2.45(13) {\AA}] and one Ge [2.52(18) {\AA}], and also three Te [3.84(18) {\AA}].\cite{Carteaux2} Experimental bond distances agree very well with each other when using different absorbing atom EXAFS spectra. The NN Te-Te distances are 3.81(90) {\AA} and 3.86(90) {\AA}, and then there are three Te-Te [4.03(6) {\AA}]. The peaks above 4.2 {\AA} are due to multiple scattering involving different near neighbours and longer bond distances. Overall, local structure-derived bond distances (Table I in \cite{Experimental}) are in agreement with bond distances from neutron powder Rietveld refinement of the average crystal structure.\cite{Carteaux2}

\textbf{Figure 3}a exhibits the temperature-dependent zero-field specific heat $C_p(T)$ of Cr$_2$Ge$_2$Te$_6$. A $\lambda$-shape peak is observed at $T_c$ = 66 K, corresponding well to the second-order PM-FM transition. The low temperature part from 2 to 10 K can be well fitted by $C_p(T) = \gamma T+ \beta T^3 + \delta T^{3/2}$ [inset in Fig. 3(a)], where the first term is the Sommerfeld electronic part, the second term is low-temperature limit of Debye phonon part, and the third term is low-temperature spin wave contribution.\cite{Gopal} The derived $\gamma$, $\beta$, and $\delta$ are 9(5) mJ mol$^{-1}$ K$^{-2}$, 2.52(4) mJ mol$^{-1}$ K$^{-4}$, and 11(3) mJ mol$^{-1}$ K$^{-5/2}$, respectively. The Debye temperature $\Theta_D$ = 198(1) K can be calculated from $\beta$ using $\Theta_D = (12\pi^4NR/5\beta)^{1/3}$, where $N = 10$ is the number of atoms per formula unit.

\textbf{Figure 3}b shows the temperature-dependent in-plane thermal conductivity $\kappa(T)$ of Cr$_2$Ge$_2$Te$_6$. In general, the $\kappa(T)$ consists of the electronic part $\kappa_e$ and the phonon term $\kappa_{ph}$, i. e., $\kappa = \kappa_e + \kappa_{ph}$. The $\kappa_e$ part can be estimated from the Wiedemann-Franz law $\kappa_e = L_0T/\rho$ with $L_0$ = 2.45 $\times$ 10$^{-8}$ W $\Omega$ K$^{-2}$ and $\rho$ is the measured electrical resistivity. The estimated $\kappa_e$ is less than 0.03 \% of $\kappa(T)$ due to the large electrical resistivity of Cr$_2$Ge$_2$Te$_6$, indicating a predominantly phonon contribution. At 300 K, the value of $\kappa(T)$ is about 5.34 W K$^{-1}$ m$^{-1}$, larger than that of 3 W K$^{-1}$ m$^{-1}$ for Cr$_2$Si$_2$Te$_6$ and, as expected due to the absence of dense grain boundaries, also larger when compared to Cr$_2$Ge$_2$Te$_6$ polycrystal.\cite{DY,Casto} In the absence of magnetic field above 100 K, the $\kappa(T)$ shows weak temperature dependence, different from phonon-dominated $\kappa$$_{ph}$($T$) calculated within Debye model (solid line in \textbf{Figure 3}b) that takes $\sim$1/$T$ behavior at high temperatures.\cite{Experimental} The difference is likely due to phonon scattering by magnetic fluctuations and strong spin-lattice coupling producing phonon glass.\cite{Casto}  With decreasing temperature, a rapid increase occurs at $T_c$ and a typical phonon peak was observed at 25 K, however stacking faults have only minor influence on $\kappa$($T$).\cite{Experimental} Strong positive field-dependent behavior of $\kappa(T)$ in \textbf{Figure 3}b is similar to RuCl$_3$ where it was described within the model of phonon scattering off Kitaev-Heisenberg excitations that are either fractionalized or incoherent type originating from strong magnetic anharmonicity.\cite{Hentrich} Since Kitaev-type magnetic interactions have been discussed in Cr$_2$Ge$_2$Te$_6$ it is likely that $\kappa$($T$) in 9 T is connected with the same mechanism.\cite{XuC1,XuC2}

\textbf{Figure 3}c presents the temperature dependence of in-plane electrical resistivity $\rho(T)$ for Cr$_2$Ge$_2$Te$_6$, showing an obvious semiconducting behavior with $\rho_{300K} = 0.45$ $\Omega$ cm. For resistivity mechanism we consider the thermally activated model $\rho(T) = \rho_0 exp(E_\rho/k_BT)$, the adiabatic small polaron hopping model $\rho(T) = AT exp(E_\rho/k_BT)$ where $k_B$ = 8.617 eV K$^{-1}$ is the Boltzmann constant and $E_\rho$ is activation energy, and the Mott variable-range hopping (VRH) model $\rho(T) = \rho_0 exp(T_0/T)^{1/4}$.\cite{Austin} \textbf{Figure 3}d shows the fitting result of the adiabatic small polaron hopping model in two temperature ranges 380-150 K and 100-60 K. The extracted activation energy $E_\rho$ is about 118.9(3) meV for 380-150 K and 26.2(7) meV for 100-60 K, respectively. Whereas the $\rho(T)$ curve can not be well fitted with the VRH model it can be explained by the thermally activated model, i.e., plot of $ln\rho$ vs $1/k_BT$ and $ln(\rho/T)$ vs $1/k_BT$ overlap with each other (\textbf{Figure 3}d).

Temperature-dependent thermopower $S(T)$ gives further insight into electronic transport. The $S(T)$ exhibits positive values in the entire temperature range of our measurement with $S_{300K} = 582$ $\mu$V K$^{-1}$ (\textbf{Figure 3}e), indicating dominant hole-type carriers. The value of $S$(T) at 300 K is quite similar to polycrystal, which argues for the absence of phonon drag effect and dominant electronic diffusion mechanism of thermal conduction.\cite{DY} The $S(T)$ from 300 to 150 K can be fitted with the equation $S(T) = (k_B/e)(\alpha+E_S/k_BT)$ (\textbf{Figure3}f),\cite{Austin} where $E_S$ is activation energy and $\alpha$ is a constant. The derived activation energy for thermopower is $E_S$ = 20(1) meV. This is much smaller than that for high temperature resistivity $E_\rho$ = 118.9(3) meV, and also smaller when compared to activation energy associated with low temperature resistivity $E_\rho$ = 26.2(7) meV. The large discrepancy between $E_S$ and $E_\rho$ typically reflects the polaron transport mechanism of carriers.

According to the polaron model, the $E_S$ is the energy required to activate the hopping of carriers, while the $E_\rho$ is the sum of the energy needed for creation of carriers and activating the hopping of carriers.\cite{Austin} Therefore, within the polaron hopping model the activation energy $E_S$ is smaller than $E_\rho$. Whereas the temperature-dependent thermopower $S(T)$ shows a change of slope anomaly at $T_c$ at 70 K, an additional change of slope is observed at 150 K, coincident with the rapid rise of electrical resistivity $\rho(T)$ on cooling through this temperature range (\textbf{Figure 3}c,e). This points to connection between electrical transport and anisotropy of 2D magnetic correlations above $T_c$, consistent with magnetic resonance experiments in low magnetic fields.\cite{Zeisner} Close inspection of $S(T)$, however, reveals an additional peak in (40 - 50) K range on cooling below $T_c$ (\textbf{Figure 3}e). To investigate this further, we turn to low-field magnetic susceptibility and X-ray scattering measurements.

\textbf{Figure 4}a,b shows the temperature dependence of dc magnetization $M(T)$ for Cr$_2$Ge$_2$Te$_6$ single crystal measured at a low field $H$ = 10 Oe applied in the $\mathbf{ab}$-plane and along the $\mathbf{c}$-axis, respectively. A PM-FM transition was observed at $T_c$ = 65 K. The $M(T)$ is nearly isotropic above, while magnetic anisotropy is observed below $T_c$. The value of $M(T)$ for $\mathbf{H\parallel c}$ is larger than that for $\mathbf{H\parallel ab}$ below $T_c$, confirming the easy $\mathbf{c}$ axis. The splitting of ZFC and FC curves at low temperature mostly originates from the anisotropic FM domain effect, in line with previous reports.\cite{Zhang,Selter,YuLiu,Liu} \textbf{Figure 4}c,d shows the temperature dependence of ZFC ac susceptibility measured with oscillated ac field of 3.8 Oe and frequency of 499 Hz. A sharp peak is observed at $T_c$ in the real part $m^\prime(T)$ for both directions. An additional anomaly occurs around 40 K just like in the $S(T)$ (\textbf{Figure 3}e), as seen a broad hump in both real part $m^\prime(T)$ and imaginary part $m^{\prime\prime}(T)$ (\textbf{Figure 4}d) due to the interplay of magnetocrystalline anisotropy and external field.\cite{Selter} The existence of anomaly below $T_c$ is confirmed by the temperature evolution of magnetic order parameter (\textbf{Figure 4}e) measured using X-ray scattering. Below $T_c$, enhanced X-ray scattering was observed at the (006) Bragg peak at a photon energy of 6 keV in the rotated, $\sigma$-to-$\pi$, scattering channel. The energy and polarization dependencies of this X-ray scattering indicate its origin as resonant magnetic scattering due to dipole transitions from the Cr 1$s$ to unoccupied 4$p$ states, which are spin-polarized by hybridization with Cr 3$d$ states. This confirms influence of magnetocrystalline anisotropy on transport behavior not only above but also below magnetic $T_c$.

Stacking faults are found in variety of 2D vdW materials such as Bi$_2$Te$_3$ or RuCl$_3$.\cite{MedlinD,YamauchiI} They originate from dislocations, core regions with large atomic displacement from the ideal position, and propagate as strain field in the crystal lattice as covalently bonded crystal units stack with vdW forces along the $c$-axis. Magnetic anisotropy in ordered state and local electronic structure may be affected by stacking faults\cite{YamauchiI,LiuYBiSe} but also polaronic defects can be created by stacking faults as a result of sub-coordinated metal atoms and electron transfer via structural sub-units.\cite{BuenoR}

Whereas surface Ge vacancies have been predicted to contribute to enhanced electrical conductivity,\cite{Song} stacking faults observed in structurally related material In$_2$Si$_2$Te$_3$ were associated with Ge(Si) pairs atomic positions that could modulate the bands near the Fermi level via Ge(Si)-Te electron transfer that indirectly alters $p$-$d$ hybridization between Cr 3$d$ and Te $p$ orbitals.\cite{Song,LefevreR} It was proposed that stacking faults lower local structure symmetry at short range distances.\cite{LefevreR} Good agreement of bond distances \textbf{Figure 2}d-f\cite{Experimental} with the average structure interatomic distances\cite{YL} imply that the stacking fault influence on the unit cell symmetry is likely at the intermediate range distances, calling for detailed total scattering analysis. Interestingly, stacking faults do not influence significantly either $\kappa$($T$) mechanism nor magnetic stripe domains that are the source of skyrmionic bubbles.\cite{Experimental,YanX,GiannoK,HanMG} Stacking faults are visible in cross-sectional Lorentz TEM sample at 13 K (\textbf{Figure 4}f). Note that dark and white lines parallel to the $c$-axis arising due to 180$^{\circ}$ stripe domains do not show significant changes upon crossing stacking faults (dark lines running perpendicular to the $c$-axis). This indicates that the faults do not perturb magnetic state or topological magnetic spin textures.\cite{HanMG} On the other hand, variation of crystal distortion induces changes in electrical transport as confirmed in an independently grown crystal.\cite{Experimental}

To qualitatively understand the potential effects of structural distortion, we calculate the band structures and conductivities in distorted (\textbf{Figure 4}h)and undistorted structures.\cite{Carteaux2,Experimental} The overall band structure of the distorted structure share a great similarity with that of undistorted structure reported previously.\cite{LeeY,MenichettiG}. \textbf{Figure 5}a,b show scalar-relativistic band structure and partial density of states (PDOS) of the distorted structure calculated with a first-principles method. An indirect gap of 0.38 eV is obtained and that is reduced by spin-orbit interaction down to 0.15 eV. The valence band maximum (VBM) is located at $\Gamma$-point in minority spin state; it is slightly off from the $\Gamma$-point in the majority spin channel. The conduction band minimums (CBM) are located between $T$ and $H_{0}$ in the majority spin channel and between $H_{0}$ and $L$ in the minority spin channel. Cr-$d$ states peak at -1 eV below Fermi level ($E_F$) in the majority spin channel and 2 eV above $E_f$ in the minority spin channel, corresponding to a spin splitting of $\sim 3$ eV. There are flat bands along the $\Gamma$--$T$ direction right below $E_F$, which may significantly contribute to transport properties. Interestingly, the positions of these flat bands are sensitive to structure distortion. \textbf{Figure 5}c,d compare the band structures along $\Gamma$--$T$ direction calculated using the distorted (crystal 1) and undisorted\cite{Carteaux2} structures. The distortion slightly shifts the flat bands BD1 and BD2 above the more dispersive band BD3 along the $\Gamma$--$T$ path. Bands BD1 and BD2 mostly consist of in-plane Te-$p_{x}$ and Te-$p_{y}$ states,\cite{Experimental} resulting in small exchange splitting and negligible dispersion along the out-of-plane direction ($\Gamma$-$T$ path). In contrast, band BD3 consists of mainly Te-$p_{z}$ and Cr-$d_{z^2}$ states, showing a much stronger dispersion along $k_z$ and larger spin splitting. Considering the proximity of the flat bands near $E_F$, their changes induced by the distortion can affect the transport properties significantly, especially at higher temperatures and with hole doping.

To explore the effects of bandstructure change on transport, we calculate the conductivities for distorted (crystal 1) and undistorted\cite{Carteaux2} structure using BoltzTraP.\cite{MadsenGK} \textbf{Figure 5}e shows the calculated $\sigma_{xx}$, the in-plane component of electrical conductivity tensor, as a function of temperature. The conductivities in both structures increase with temperature, as expected for intrinsic semiconductors. The distorted structure has a higher in-plane conductivity than the undistorted structure. The important contribution of BD1 and BD2 on conductivity can be further illustrated by calculating the predicted conductivities in the hole-doped case, as shown in \textbf{Figure 5}f. The chemical potential $\mu_2$ is lowered to 0.05 eV below $E_f$, as denoted in \textbf{Figure 5}c by green dash-doted line, to mimic the hole-doping effects. As is expected, the conductivities of all cases increase at lower temperatures. The $\sigma_{xx}$ of distorted structure, however, is increased more than 100 times. Therefore, first-principle calculations indicate that the local distortion with appropriate doping can profoundly affect electronic and magnetic properties of Cr$_2$Ge$_2$Te$_6$, leading to ever higher conductivity.

\section{CONCLUSIONS}

In summary, our results indicate strong influence of crystal structure distortion and magnetocrystalline anisotropy on electrical and thermal transport both above and below $T_c$ in 2D vdW magnetic Cr$_2$Ge$_2$Te$_6$ crystals. Electronic transport is dominated by polaronic effects commonly observed in oxide materials, confirming strong electron-phonon coupling. Moreover, interplay between spin-orbit electron-phonon coupling may tailor the spin polarization since the polaron could retain only one of the spin-polarized bands in its coherent spectrum.\cite{Covaci} This could be used in spin-orbitronic and magneto-thermal devices that exploit current propagation via spin-orbit torque mechanism.\cite{Watanabe,Ostwal}

\section{Methods}

Crystals were synthesized from the self-flux method \cite{Experimental}. Crystals were cut in suitable geometry for in-plane resistivity measurements. Pulverized crystals were used XANES and EXAFS measaurement. Reference\cite{Experimental} describes experimental methods in more details.

\section*{Acknowledgements}

Work at Brookhaven National Laboratory (BNL) is supported by the Office of Basic Energy Sciences, Materials Sciences and Engineering Division, U.S. Department of Energy under Contract No. DE-SC0012704. TEM sample preparation was carried out at the Center for Functional Nanomaterials (BNL), which is supported by the US Department of Energy, Office of Basic Energy Sciences under Contract No. DE-AC02-98CH10886. This research used beamlines 4-ID and 8-ID of the National Synchrotron Light Source II, a U.S. Department of Energy (DOE) Office of Science User Facility operated for the DOE Office of Science by Brookhaven National Laboratory under Contract No. DE-SC0012704. The work carried out at the University of Delaware was supported by the U.S. Department of Energy, Office of Science, Basic Energy Sciences, under Award DE-SC0008885. First principle calculations were supported by the U.S.~Department of Energy, Office of Science, Office of Basic Energy Sciences, Materials Sciences and Engineering Division. Ames Laboratory is operated for the U.S. Department of Energy by Iowa State University under Contract No.~DE-AC02-07CH11358. A portion of this work was performed at the National High Magnetic Field Laboratory, which is supported by the NSF Cooperative Agreement No. DMR-1644779 and the State of Florida.

\begin{figure*}
\centerline{\includegraphics[scale=0.38]{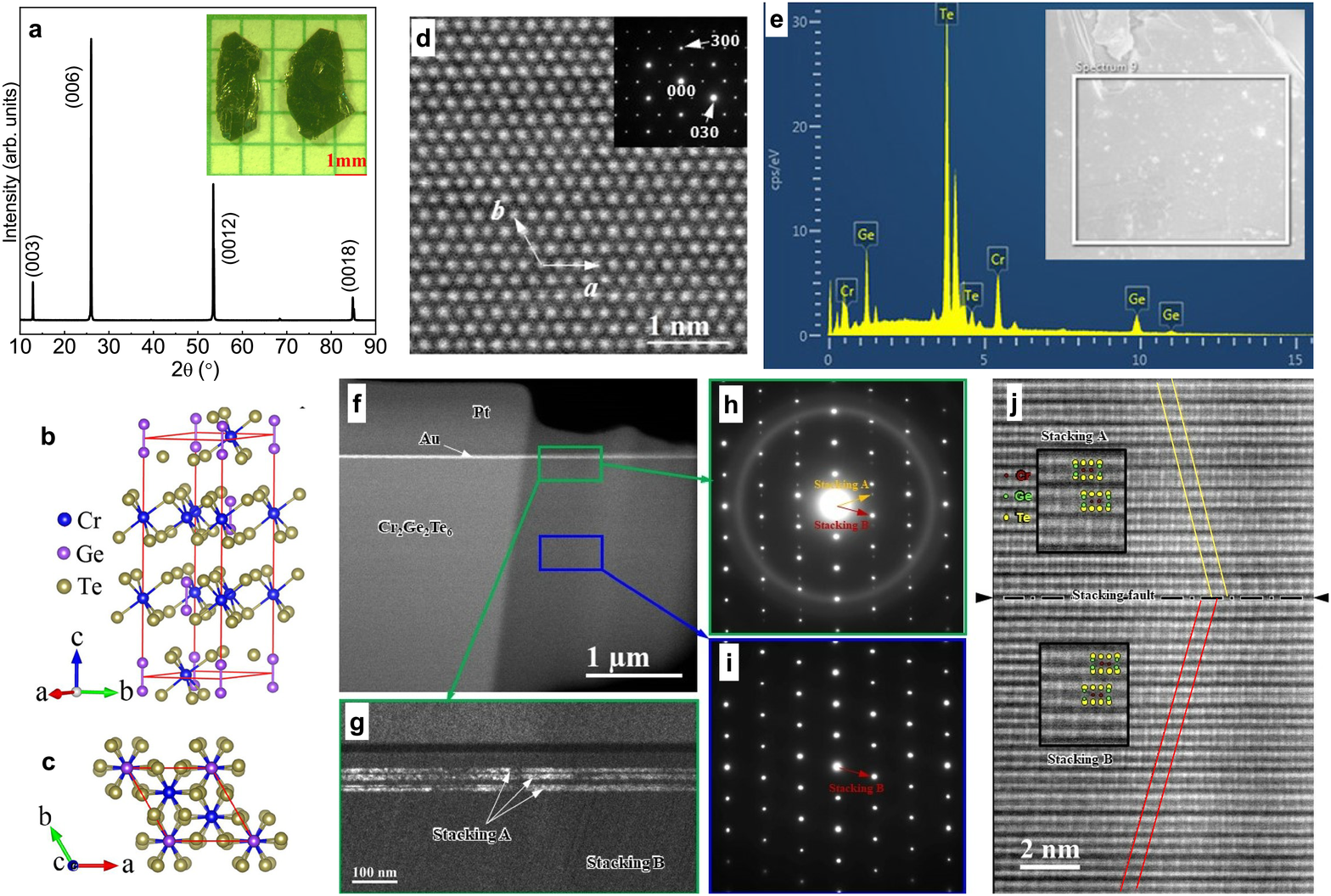}}
\caption{Crystal structure and nanostructure characterization of Cr$_2$Ge$_2$Te$_6$. a) 2$\theta$ X-ray diffraction scan along the (00l) direction. Inset shows optical image of Cr$_2$Ge$_2$Te$_6$ single crystals. $R\bar{3}h$ unit cell of Cr$_2$Ge$_2$Te$_6$ shown from b) the side view and c) the top view. d) High-angle annular dark-field (HAADF) scanning transmission electron microscopy (STEM) image obtained along the $\mathbf{c}$-axis. Inset shows the electron diffraction pattern. e) X-ray energy-dispersive spectrum of the selected area. HAADF STEM image f) of cross-sectional TEM sample imaged along the $\mathbf{a}$-axis. Dark-field TEM image g) obtained with the reflection of stacking A (orange arrowed peak in h). Selected area electron diffractions (h and i) from the green-boxed area (h) and blue-boxed area (i). In (i) the Stacking A only exists. Atomic resolution HAADF STEM image j) from the green-boxed area. Two insets show the enlarged area showing the two different stackings separated by a stacking fault, indicated with a dotted line. Two inclined yellow and red lines show the stacking sequence of the Te/Ge columns along the $\mathbf{c}$-axis.}
\label{res}
\end{figure*}

\begin{figure*}
\centerline{\includegraphics[scale=1.0]{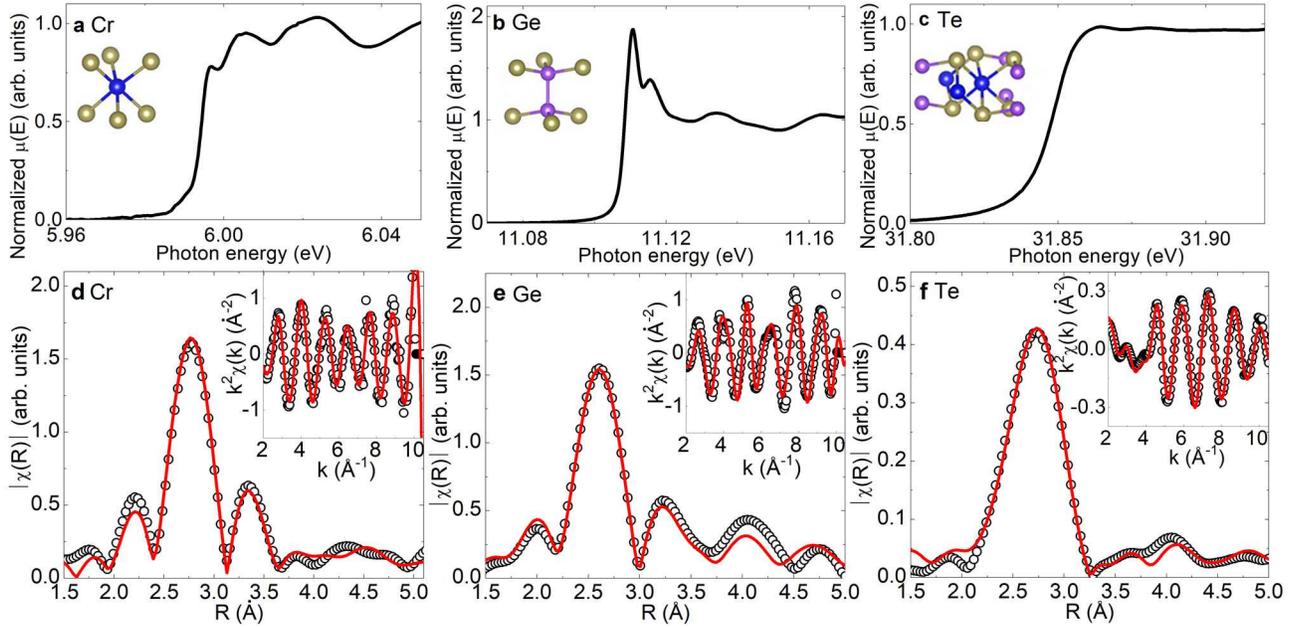}}
\caption{Local atomic structure. Normalized a) Cr, b) Ge, and c) Te K-edge XANES spectra with insets depicting local atomic structure. Fourier transform magnitudes of the EXAFS oscillations (symbols) for d) Cr, e) Ge, and f) Te K-edge with the phase shifts correction. The model fits are shown as solid lines. Insets show the corresponding filtered EXAFS (symbols) with $k$-space model fits (solid lines).}
\label{exafs}
\end{figure*}

\begin{figure*}
\centerline{\includegraphics[scale=0.5]{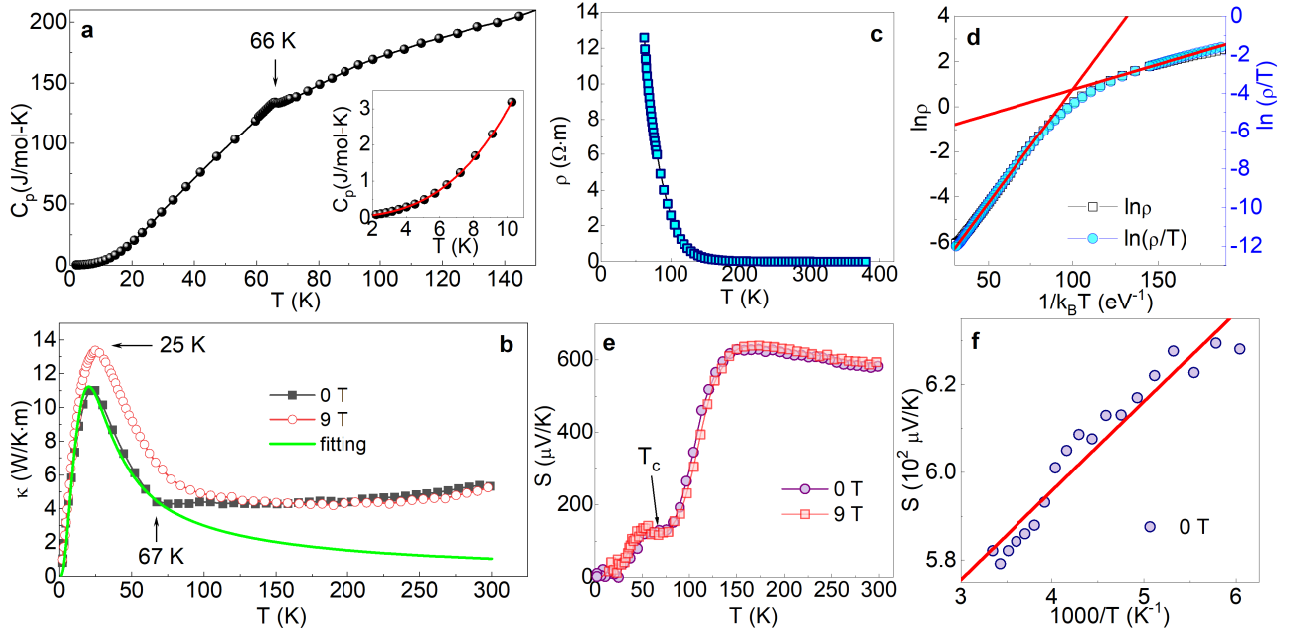}}
\caption{Thermodynamic, electronic and thermal transport properties. Temperature dependence of a) specific heat $C_p(T)$, b) thermal conductivity $\kappa$($T$) and c) electrical resistivity $\rho(T)$ of our Cr$_2$Ge$_2$Te$_6$ single crystals. Inset in a) shows the fitted (red line) low-temperature $C_p(T)$ from 2 to 10 K (see text). Green solid curve in b) shows the fitting of the lattice thermal conductivity using the Debye model. The ln($\rho$) vs $1/k_BT$ (left axis) and ln($\rho/T$) vs $1/k_BT$ (right axis) d) fitted by adiabatic small polaron hopping model (see text). e) Temperature dependence of $S(T)$thermopower. f) The $S(T)$ vs, $1000/T$ curve fitted by polaron transport model (see text).
}
\label{exafs}
\end{figure*}

\begin{figure*}
\centerline{\includegraphics[scale=0.38]{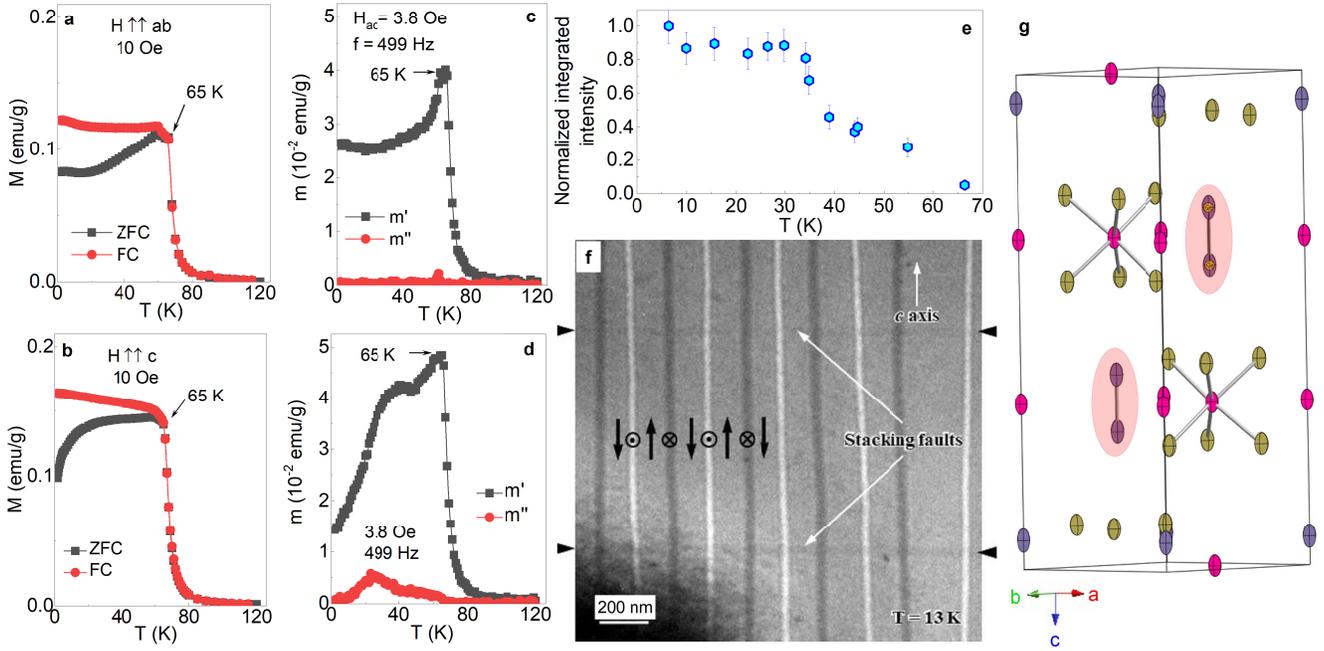}}
\caption{Magnetocrystalline anisotropy. Temperature dependence of zero field cooling (ZFC) and field cooling (FC) dc magnetization $M(T)$ measured at H = 10 Oe for Cr$_2$Ge$_2$Te$_6$ single crystal with a) $\mathbf{H\parallel ab}$ and b) $\mathbf{H\parallel c}$, respectively. Temperature dependence of ac susceptibility real part $m^\prime(T)$ and imaginary part $m^{\prime\prime}(T)$ measured with oscillated ac field of 3.8 Oe and frequency of 499 Hz applied c) in the $\mathbf{ab}$-plane and d) along the $\mathbf{c}$-axis, respectively. e) The integrated intensities of longitudinal scans through the (006) magnetic Bragg peak of Cr$_2$Ge$_2$Te$_6$. f) Lorentz TEM image of Cr$_2$Ge$_2$Te$_6$ obtained perpendicular to the $\mathbf{c}$-axis. at 13 K, showing 180$^\circ$ stripe domains separated by the Bloch walls due to uniaxial anisotropy (Ref. 61). Two stacking faults are visible as dark horizontal lines, which do not affect the stripe domain configuration. g) Unit cell of distorted Cr$_2$Ge$_2$Te$_6$ showing thermal ellipsoids from single crystal refinement and Ge-Ge bonds (red areas).}
\label{res}
\end{figure*}

\begin{figure*}
\centerline{\includegraphics[scale=0.4]{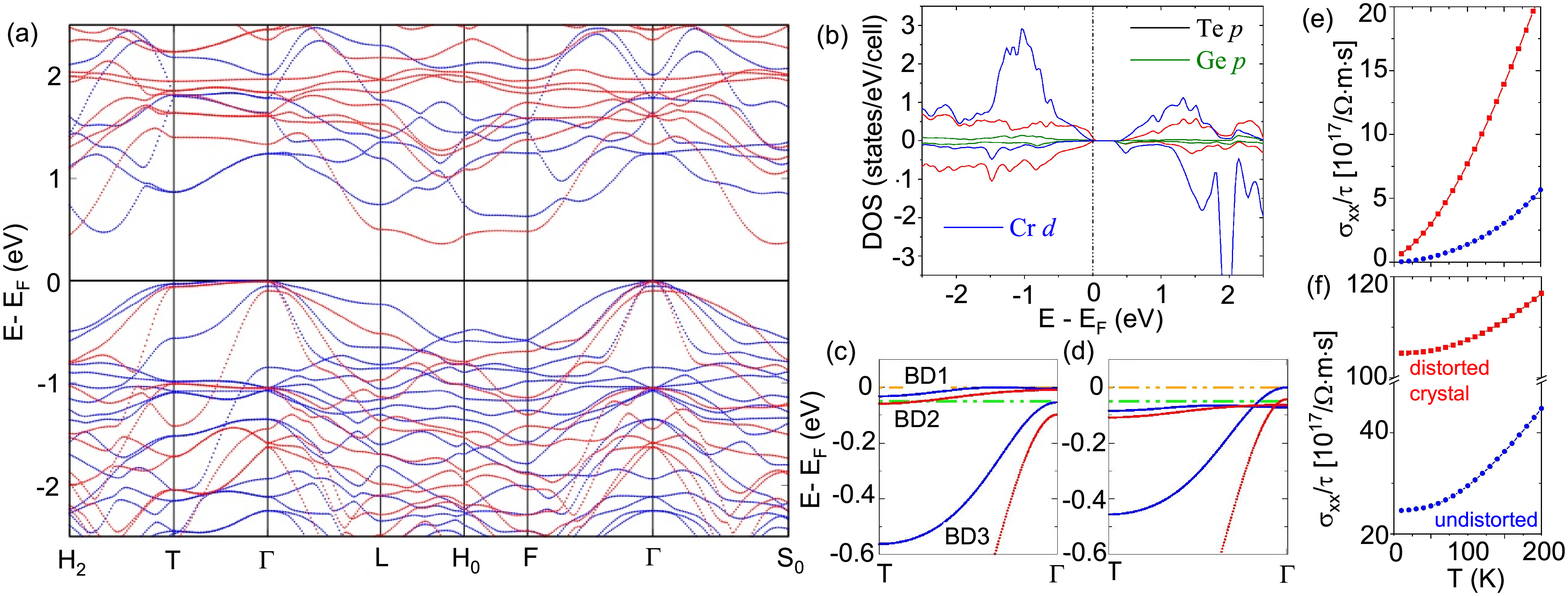}}
\caption{(a) Scalar-relativistic band structure of Cr$_2$Ge$_2$Te$_6$ with lattice distortion. Blue (red) color indicates the majority (minority) spin band. (b) The corresponding partial density of states (PDOS) projected on Te-$p$, Ge-$p$, and Cr-$d$ states. Scalar-relativistic band structures calculated in (c) distorted and (d) undistorted structures. BD1, BD2, and BD3 are selected bands to analysis band characters (See Ref. 40). Orange and green dash-dotted lines denote two chemical potentials, $\mu$1 and $\mu$2, respectively, that are employed for conductivity calculations. Electrical conductivity as a function of temperature calculated without (e) and with (f) hole-doping in Cr$_2$Ge$_2$Te$_6$ distorted and undistorted structures. The chemical potential is set to 0.05 eV below VBM to mimic the hole-doped case. The red and blue colors denote the distorted and undistorted structures, respectively.}
\label{res}
\end{figure*}

\end{document}